# Progress in Analytical Solutions for High Order Harmonic Generation in Semiconductor Superlattice Multipliers

Abdullah Al-Ateqi[1] and Mauro Fernandes Pereira[1,2*]

[1]Department of Physics, Khalifa University of Science and Technology, Abu Dhabi, 127788, UAE.
[2]Institute of Physics, Czech Academy of Sciences, Prague, 18221, Czech Republic.

*Corresponding author(s). E-mail(s): mauro.pereira@ku.ac.ae;
Contributing authors: 100053593@ku.ac.ae;

**Abstract**

In this study, we address the limitations of previous solutions for modeling high-order harmonics in semiconductor superlattices (SSLs). Earlier research proposed a step function ansatz that effectively modeled high-order even and odd harmonics but introduced numerical noise. An upgrade using a logistic function addressed the noise problem but eliminated high-order odd harmonics. To overcome both limitations, we examined the impact of the y-intercept value in the discontinuity of the step function and proposed a modified logistic function as a new ansatz. The modified logistic function delivers accurate results, preserving high-order odd harmonics up to the 50$^{th}$ order similar to the previous ansatz while also eliminating numerical noise. This research contributes to a more efficient and robust analytical approach for modeling SSLs, notably by avoiding time-consuming numerical solutions and enhancing our understanding of nonlinear phenomena in GHz-THz devices.

# 1 Introduction

There are many applications in the subTHz-THz-MIR range and advanced materials are in demand [1–8]. Semiconductors [1, 9–24], specially semiconductor superlattices (SSLs) allow a controlled study of both sources and detectors [25–39]. In materials science, a superlattice means periodic arrangement of alternating materials. However, in photonics, a waveguide superlattice is typically an array of waveguides made of a single material. Recent achievements of photonics superlattices in the NIR can be modified to address novel functionalities in the THz-MIR range [40–43]. We should, of course, mention quantum cascade lasers and detectors, which just like superlattices are a very competitive technology, albeit expensive, for the THz-MIR range [44–51].

One of the reasons that SSLs have emerged as a promising nanomaterial for a wide range of applications in the Sub-Terahertz and Terahertz (THz) region is their ability to generate high-order harmonic generation (HHG) [52]. These structures play a crucial role in controllable nonlinear optical effects, such as HHG, reaching up to the 54$^{th}$ harmonic at 37$\mu m$ [53–56].

Note further that the lower frequency of the up to the THz spectrum, which extends fully from 0.1 to 10 THz, can be reached through frequency multiplication of input synthesizers, such as Backward Wave Oscillators (BWO's), or widely commercially available Schottky diodes.

However, for efficient operation in the THz range, the cut-off frequency of Schottky diodes must be increased, which is difficult due to fundamental restrictions, such as high inertness of electron transit through the active zone and parasitic capacitances. In contrast, our choice of device, i.e. Semiconductor Superlattice Multipliers. (SSLMs), can overcome these difficulties as multipliers due to their intrinsic low values of both response time and parasitic capacitances, coupled with substantial negative differential conductivity (up to 1 THz) in the current–voltage curves. [53]. This range of frequencies is becoming increasingly important for metabolomics applications, where blood, urine, and breath can be analyzed spectroscopically as well as for studies of energetic materials [29, 57–59]. Furthermore, a recent review with a perspective analysis of Urinary Metabolic Biomarker Profiling for Cancer Diagnosis by Terahertz Spectroscopy: Review and Perspective is given in Ref. [60].

The fundamental mechanism of SSL multiplication involves electrons accelerated under a static field with amplitude $E_{dc}$, resulting in Bloch oscillations. These oscillations are responsible for the robust nonlinear current-voltage characteristic of SSLs and are modulated by an oscillating electromagnetic field, giving rise to HHG [61]. Both electric field domains and excitonic effects can significantly impact this process. Previous studies used a Nonequilibrium Green's Functions (NEGF) approach to analyze HHG in biased superlattices when an oscillating field modulates the Bloch oscillations. The approach allowed prediction and measurement of the control of even harmonics by orders of magnitude, using a miniband transport model for the Boltzmann equation within the relaxation rate approximation. To analyze the results, the superlattice semiconductor model (SSLM) is exposed to a combined static bias and a monochromatic field, *E=$E_{dc}$+$E_{ac}$cos (2πvt)*. This field leads to an expression for the electromagnetic power emitted by the nth harmonic. The parameter *α = e$E_{cd}$d/ hv* controls the nonlinear response of the system [25]. The rectified potential drop, $U=eE_{cd}d+mhv$, affects the static I–V, leading to interesting current rectification confirmed by experiments [62]. The dephasing $\Gamma=k/\tau=U_c$ can be calculated by NEGF [27] or extracted from static current-voltage measurements [27]. The measured I–V curves deviate from the ideal Ezaki-Tsu model due to factors such as the current density flows in different directions under positive or negative bias and interface roughness. The effect of interface roughness has been successfully modeled, providing excellent comparisons between theory and experiments [27]. Note that the NEGF approach delivers precise I-V to predict and understand first-order physical processes in semiconductor superlattices. Our theory and simulations are focused on the nonlinear, high-order case, where NEFG codes do not operate well for high-intensity and low-frequency input considered here. Thus, an ansatz was introduced, which described this asymmetric flow effect [27]. However, this led to numerical instabilities for low incident powers, making it hard to distinguish from experimental noise. Thus, a new solution that is fully differentiable at zero was introduced to solve this issue [34], but it resulted in the loss of the odd harmonics at high order, despite being experimentally observed. Thus, this study aims to develop an improved ansatz for better prediction of harmonic power in SSLMs. At this point, we should point out that control of anisotropy in surfaces, which have some similar characteristics with the interfaces with intrinsic anisotropy is of large importance in III-V structures with magnetic layers [63, 64]. Likewise, surface-nanoengineered III-V Schottky diodes, which are also relevant for frequency multiplication, play an equally relevant role in the detection, notably of amino-acids [65, 66], which parallels the use of SSLMs in the detection of nitriles [29].

## 2 Methods/Description

The analytical ansatz solution was previously utilized to describe asymmetric current flow and the effects of asymmetric scattering on HHG in the context of superlattice semiconductors (SSL) [27, 30, 31, 33]. Considering an SSL with a period d under an electric field $E_{dc}+E_{ac}cos(2\pi vt)$, the time-dependence of the current response can be represented by the Fourier basis:

$$j^v(t)=j_{dc}^v+\sum_{l=1}^{\infty}\left[j_l^{v,cos}cos(2\pi vlt)+j_l^{v,sin}sin(2\pi vlt)\right], (1)$$

$$j_{dc}^v=\sum_{l=1}^{\infty}J_n^2(\alpha)j_{dc}(U), (2)$$

$$j_l^{v,cos}=\sum_{n=-\infty}^{\infty}J_n^2(\alpha)\left[J_{l+n}^2(\alpha)+J_{l-n}^2(\alpha)\right]j_{dc}(U), (3)$$

$$j_l^{v,sin}=\sum_{n=-\infty}^{\infty}J_n^2(\alpha)\left[J_{l+n}^2(\alpha)-J_{l-n}^2(\alpha)\right]K(U), (4)$$

here the dc current ($j_{dc}^v$) is given by Eq. (2) and the Fourier components $j_l^{v,cos}$, $j_l^{v,sin}$ describe the lth harmonic generation. $J_l$ is the Bessel function of the first kind of order *l*, and *U=u+phv*, where $u = eE_{dc}d$ is the energy drop per period due to the dc bias. The critical parameter $\alpha = E_{dc}d/h\nu$ appears in (2),(3),(4). Combining these parameters with the relaxation approximation, $j_{dc}(U)$ and *K (U)* can be expressed in a simple format:

$$j_{dc}(U)=2j_0 \frac{U/\Gamma}{1+(U/\Gamma)^2}, \ K(U)=\frac{2j_0}{1+(U/\Gamma)^2}, \ (5)$$

Here, $j_0$ is the peak current corresponding to the electric field $E_c=\hbar/ed\tau$, and $\Gamma$ is the scattering rate. However, in real SSL, the interface of GaAs over $Al_{1-x}Ga_x$ As is worse than $Al_{1-x}Ga_x$ As over GaAs due to surface imperfection, which makes the flow from right to left unequal to that from left to right [30]. Therefore, an ansatz *f(x)* was used to modify the Tsu and Esaki limit, such that [67]

$$j_0=j_0^- + (j_0^+ - j_0^-)f(x), \ \Gamma=\Gamma^- + (\Gamma^+ - \Gamma^-)f(x), \ (6)$$

$$f(x)=F(x)=\begin{cases} 1 & \text{if } x>0 \\ 0 & \text{if } x<0 \end{cases}, \ (7)$$

where $x=U/\Gamma^*$. The step function ansatz *F(x)* succeeded in explaining the experimentally detected even harmonic generation but led to numerical errors for low incident powers due to the discontinuity at the origin. Another ansatz, the logistic function *L(x)* was introduced to address this issue:

$$f(x)=L(x)=\frac{1}{(1+e^{-kx})}, \ (8)$$

This ansatz removed the discontinuity at zero, and due to its differentiable nature, solved the numerical error. However, it raised another issue, that is, eliminating high-order odd harmonics [34]. Leading the study to consider another possibility, combining the properties of both the step function and the logistic function. In particular, the value at the y-intercept of the step function *F(0)=1*, in addition to the smooth increase of the logistic function around the y-intercept, both were necessary to maintain the high-order odd harmonics and to reduce the numerical noise, respectfully. To achieve this, we propose the following modification of the logistic function, we called the modified logistic function *S(x)*:

$$f(x)=S(x)=\begin{cases} 1 & \text{if } x=0 \\ \frac{1}{(1+e^{-kx})} & \text{if otherwise} \end{cases}, \ (9)$$

Before we proceed, it is important to highlight that the harmonic power generated by electron oscillation induced by the nonlinear current (actually, modulated Bloch oscillations) is given by

$$P_n=I\left[\left(j_n^{v,cos}\right)^2+\left(j_n^{v,sin}\right)^2\right], \ (10)$$

The constant $P_n$ is derived in detail in Ref. [68] Furthermore, we should note that there is an analytical exact form for the step function, [69] would look ideal to implement, namely

$$F(x)=\frac{1}{\pi}\left(\frac{3\pi}{4}+\arctan(x-1)+\arctan\left(\frac{2-x}{x}\right)\right), \quad (11)$$

However, the numerical analysis in the next section shows that it is not the best approximation compared to the modified sigmoid approximation proposed here. The current flow asymmetry can be attributed to the difference in interface quality between GaAs/AlAs and AlAs/GaAs. This can be smoothly described by our NEGF method, through an interface scattering selfenergy [27], or by our path integral method [32]. The Ansatze used here are efficient way to reproduce these curves and Eq. (9) gets closer to the direct numerical calculations, delivering both the harmonic generation at high orders and the low voltage, low power studies with smooth curves, further explaining the asymmetric output, seen experimentally [61].

## 3 Numerical Results and Discussion

The following numerical results are for *GaAs- AlAs* SSLs with a period of *d* = 6.23*nm*, subject to an oscillating field of 141*GHz*. The primary input parameters for Eq. (1) (2)(3)(4)(5)(6)(7)(8) are $j_0^+$ and $j_0^-$ equal to 2.14 and 1.94×$10^9$*A/$m^2$*, respectively; $\Gamma^+$ and $\Gamma^-$ equal to 21 and 20×$10^9$*meV*, respectively. Note that a detailed experimental analysis led to the experimental detection up to harmonic 54 [53]. Since we need to determine the validity of our modified logistic function approach for high order harmonics, Figs. 1-4 cover a wide harmonic range.

Figure 1 illustrates that the y-intercept *f(0)*=1/2 almost fully symmetrizes the I-V for high harmonics, eliminating the odd harmonics, while the *f(0)*=0 and *f(0)*=1 equally preserve the output. The three cases summarize the possible options of symmetrizing the step function at the origin or not.

Intuitively, as *k* increases *k*→∞, the modified logistic function becomes indistinguishable from the regular step function with a y-intercept of 1. Thus, giving the same results. Note however that *k*=30 marks the limit value for the use of the modified logistic Ansatz before numerical errors develop in the study of the bias dependence, which we investigate here since this kind of study has been performed experimentally [61]. As in our previous work [34], the asymmetry in the experimental data (difference in power peaks from left to write), which could not be explained by the symmetric flow model used in Ref. [61], can be attributed to the asymmetric current flow (*δ*≠1).

The modified logistic ansatz successfully addressed both the vanishing high-order odd harmonics issue introduced by the logistic function and the numerical error at low power introduced by the step function. As demonstrated in Figs. 2 and 3, the power of odd harmonics for the modified logistic ansatz is consistent with those of the step function, as it resolves the extremely low values of high-order odd harmonics caused by the logistic function. Furthermore, the modified logistic function exhibits error-free plots for low power at *α = 4* and 8. Both problems are resolved by combining the smooth increase from *f(x)=0* to *f(x)=1*, which maintains an error-free power plot, with a *f(0)=1*, akin to the step function, which prevents the vanishing of high-order odd harmonics.

Although the modified logistic function ansatz effectively resolves both issues arising from the previous ansatz, it can be observed that the even harmonics diminish more rapidly compared to the step function ansatz. This observation is attributed to the smoothness of the increase from 0 to 1 ruled by the $k$ parameter in Eq. (9), as depicted in Fig. 4. For instance, when using $k$=80, which results in a more vertical increase, the calculated harmonics power increases and aligns more closely with the step function ansatz. However, this vertical increase also introduces errors in the harmonic power calculations, as evidenced by small noise spikes observed in Fig. 3. Consequently, it can be concluded that a smoother curve (lower $k$ value) results in reduced calculation noise. Therefore, the $k$ value can be adjusted to optimally match even harmonics experimental values while maintaining the lowest possible $k$ value to minimize error.

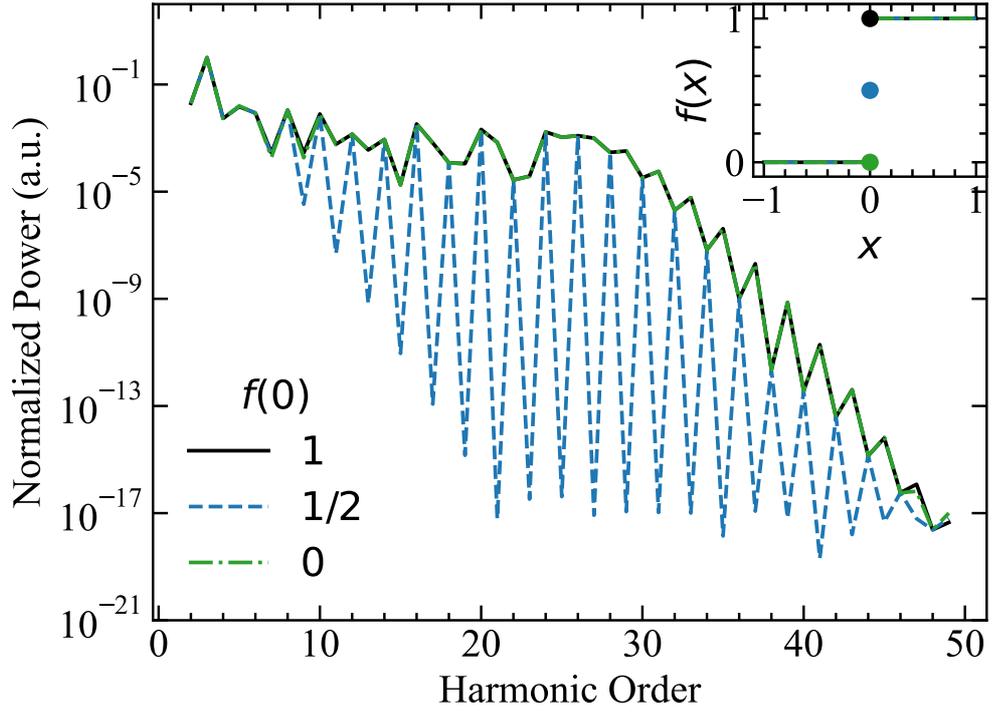

**Fig. 1** Results of the calculated normalized emitted power for the corresponding step function ansatz with varying values of *f(0)=0*, *f(0)=1/2*, and *f(0)=1*, for the $n^{th}$ harmonic spanning the range from the $2^{nd}$ to the $50^{th}$ harmonic, for an SSL subject to an incident power of ≈*47μW* (equivalent to $\alpha = 28.3$), and an oscillating field of *141GHz*.

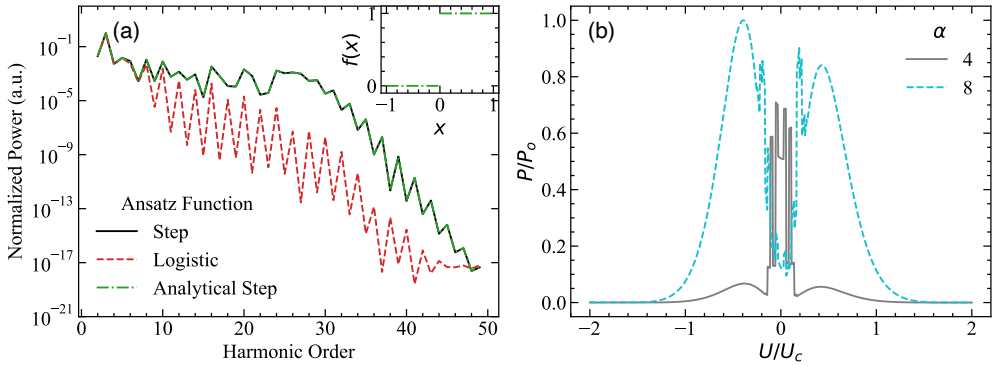

**Fig. 2** Comparison and contrast of the emitted harmonic power using the Step, Logistic, and analytical step ansatz functions. **a,** Results of the normalized emitted harmonic power for harmonics ranging from the $2^{nd}$ to the $50^{th}$, for the $n^{th}$ harmonic, for an SSL subject to an incident power of ≈*47μW* (equivalent to $\alpha = 28.3$), and an oscillating field of *141GHz*.. The inset illustrates the analytical step function curve. **b,** The power output of the second harmonic for an SSL subject to an incident power $\alpha = 4$ and 8, using the analytical step function ansatz.

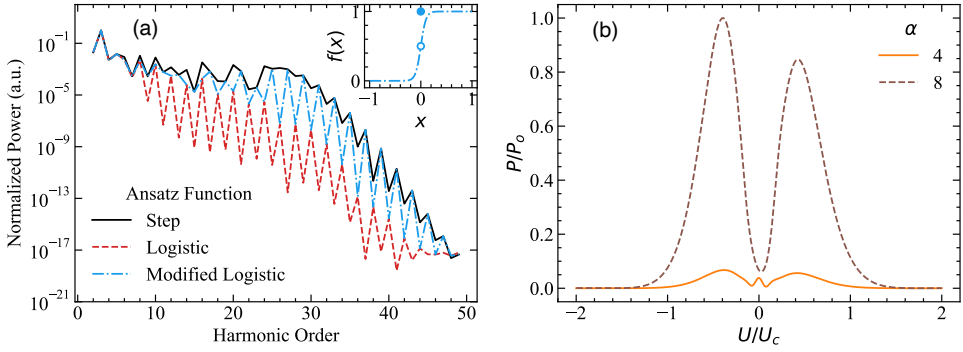

**Fig. 3** Comparison and contrast the emitted harmonic power using the Step, Logistic, and Modified Logistic ansatz functions. **a,** Results of the normalized emitted harmonic power for harmonics ranging from the 2$^{nd}$ to the 50$^{th}$, for the $n^{th}$ harmonic, for an SSL subject to an incident power of ≈47$\mu W$ (equivalent to $\alpha$ = 28.3), and an oscillating field of 141$GHz$. The inset illustrates the modified logisticfunction curve. **b,** The power output of the second harmonic for an SSL subject to an incident power $\alpha$ = 4 and 8, using the modified logistic function ansatz with $k$=20.

Furthermore, a detailed analysis shows that the current flows asymmetry, ruled by the current flow asymmetry parameter $\delta = \Gamma^+/\Gamma^- = j^+/j^-$, used [31, 33, 34, 68, 70], increases the power output of all harmonics, notably through a strong enhancement of the $j_l^{\nu,sin}$ contribution in Eq. (4). The $f(0) = 1/2$ in representations of the delta function, such as the logistic function, strongly reduces this term and eliminates the high-order harmonics.

As a final test, we compare and contrast the modified logistic ansatz with previous experimental data in Fig. 5.

The new modified logistic function ansatz model shows strong agreement with the experimental values, demonstrating its effectiveness in representing the relationship between normalized power and harmonic order. The impact of different $k$ values before the 7$^{th}$ harmonic is minimal, indicating that choosing a $k$ value of 20 or 30 is sufficient for accurate noise-free modeling of the power vs. harmonic order relationship.

# 4 Conclusion

In conclusion, this paper systematically investigated the limitations of previous ansatz solutions for modeling high-order harmonics in semiconductor superlattices (SSLs) and proposed a modification of the ansatz to overcome these limitations. This modification addresses two main issues: the vanishing of high-order odd harmonics and numerical noise that develops in bias-dependent investigations.

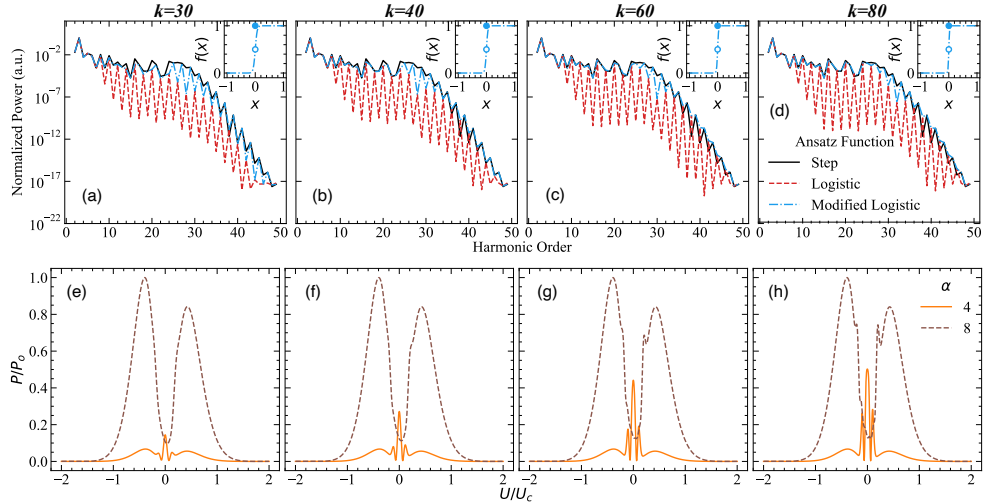

**Fig. 4 Extended plots for the emitted power results using the ansatze with *k* = 30, 40, 60, and 80. (a-d),** The normalized emitted power for the *n*th harmonic shows the extended plot for harmonics ranging from the 2nd to the 50th, for an SSL subject to an incident power of ≈47μW (equivalent to *α = 28.3*), and an oscillating field of *141GHz*. **(e-h),** Results of the corresponding power output of the second harmonic for an SSL subject to incident power *α* = 4 and 8, using the modified logistic function ansatz with *k* = 20.

The analysis focused on the impact of the y-intercept value in the step function ansatz and demonstrated that the selection of the y-intercept point value is crucial in the magnitude of the $j_l^{v,sin}$ term. Moreover, the new results reveal that the smooth increase from *f(x)=0* to *f(x)=1* ruled by the value of *k* plays a strong role in the reduction of the calculation noise. A low *k* value results in lower calculation noise, and lower even harmonics power output while a higher *k* value results in noise spikes at low input power while yielding harmonic power values closer to those obtained from the step function ansatz. The modified logistic function ansatz, with the appropriate value of *k*, successfully addresses both limitations up to the 50th order. By carefully selecting the value of *k*, it is possible to balance the trade-off between noise reduction and the preservation of high-order even harmonics.

Compared with the available experimental power values which are up to the 7th harmonic, *k* values of 20 and 30 demonstrated a strong agreement with experimental values. This proves that the modified logistic ansatz with these *k* values can be sufficient to provide an accurate and noise-free model for low-order harmonic generation.

The introduction of the modified logistic function ansatz, along with a deeper understanding of the role of the y-intercept and of its control parameter, has led to a more efficient and robust analytical approach for modeling SSLs. This research advanced our knowledge of nonlinear phenomena in GHz-THz devices and has the potential to support the development of more efficient applications of SSLM-based spectroscopy systems.

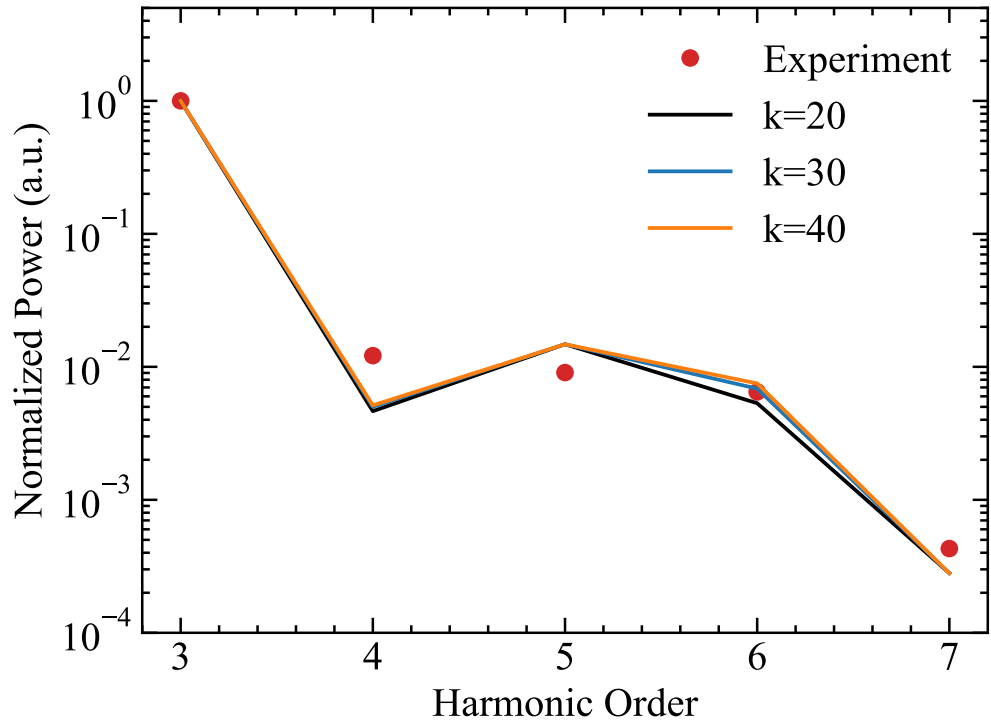

**Fig. 5 Theoretical and experimental normalized emitted power.** The theoretical values were computed using the modified logistic ansatz with *k* = 20, 30, and 40, for an SSL subject to an incident power of ≈*47μW* (equivalent to *α* = 28.3), and an oscillating field of 141*GHz*. The experimental data [27] shown in the plot were calculated by removing the waveguide effects, for a complete comparison.

# Declarations

## Ethical approval

Not applicable.

## Competing interests

The authors declare no competing interests.

## Authors' contributions

M.F.P. obtained funding, provided input equations, computer codes and research coordination; A.A. proposed the modified logistic Ansatz and implemented it; All authors contributed to the manuscript preparation.

## Funding

This publication is based upon work supported by Khalifa University of Science and Technology under Award No. CIRA 2021-108.This publication is based upon work supported by Khalifa University of Science and Technology under Award No. CIRA 2021-108.

## Availability of data and materials

All simulations presented in this paper were obtained with the author's computer codes. The data files used to generate the Figures can be obtained from the authors upon reasonable request.

## References

[1] Radovanović, J., Milanović, V., Ikonić, Z., Indjin, D., Harrison, P.: Electron-phonon relaxation rates and optical gain in a quantum cascade laser in a magnetic field. Journal of Applied Physics **97**(10), 103109 (2005)

[2] Shields, T., Dada, A.C., Hirsch, L., Yoon, S., Weaver, J.M.R., Faccio, D., Caspani, L., Peccianti, M., Clerici, M.: Electro-optical sampling of single-cycle thz fields with single-photon detectors. Sensors **22**(23) (2022)

[3] Ng, R.A., Portnoi, M.E., Hartmann, R.R.: Tuning terahertz transitions in cyclo[*n*]carbon rings. Phys. Rev. B **106**, 041403 (2022)

[4] Hartmann, R.R., Portnoi, M.E.: Guided modes and terahertz transitions for two-dimensional dirac fermions in a smooth double-well potential. Phys. Rev. A **102**, 052229 (2020)

[5] Saroka, V.A., Hartmann, R.R., Portnoi, M.E.: Terahertz transitions in narrow-gap carbon nanotubes and graphene nanoribbons. Journal of Physics: Conference Series **1092**(1), 012121 (2018)

[6] Collier, T.P., Saroka, V.A., Portnoi, M.E.: Tuning terahertz transitions in a double-gated quantum ring. Phys. Rev. B **96**, 235430 (2017)

[7] Collier, T.P., Saroka, V.A., Portnoi, M.E.: TUNING THz TRANSITIONS IN A QUANTUM RING WITH TWO GATES, pp. 172–175. https://doi.org/10.1142/9789813224537 0040

[8] Pereira, M.F., Shulika, O.: Terahertz and Mid Infrared Radiation: Detection of Explosives and CBRN (using Terahertz). Springer, ??? (2014)

[9] Radovanović, J., Milanović, V., Ikonić, Z., Indjin, D.: Quantum well shape optimization of continuously graded al$_x$ga$_{1-x}$N structures by combined super-symmetric and coordinate transform methods. Phys. Rev. B **69**, 115311 (2004)

[10] Ilić, I., Beličev, P.P., Milanović, V., Radovanović, J.: Analysis of tunneling times in absorptive and dispersive media. J. Opt. Soc. Am. B **25**(11), 1800–1804 (2008)

[11] Isić, G., Radovanović, J., Milanović, V.: Anisotropic spin-dependent electron tunneling in a triple-barrier resonant tunneling diode. Journal of Applied Physics **102**(12), 123704 (2007)

[12] Radovanović, J., Milanović, V., Indjin, D., Ikonić, Z.: Intersubband absorption at 1.3m in optimized GaN/AlGaN Bragg-confined structures. Journal of Applied Physics **92**(12), 7672–7674 (2002)

[13] Gajic, R., Kuchar, F., Meisels, R., Radovanovic, J., Hingerl, K., Zarbakhsh, J., Stampfl, J., Woesz, A.: Physical and materials aspects of photonic crystals for microwaves and millimetre waves. International Journal of Materials Research **95**(7), 618–623 (2004) https://doi.org/10.1515/ijmr-2004-0118

[14] Isić, G., Indjin, D., Milanović, V., Radovanović, J., Ikonić, Z., Harrison, P.: Phase-breaking effects in double-barrier resonant tunneling diodes with spin-orbit interaction. Journal of Applied Physics **108**(4), 044506 (2010)

[15] Milanović, V., Radovanović, J., Ramović, S.: Influence of nonparabolicity on boundary conditions in semiconductor quantum wells. Physics Letters A **373**(34), 3071–3074 (2009)

[16] Isić, G., Milanović, V., Radovanović, J., Ikonić, Z., Indjin, D., Harrison, P.: Time delay in thin slabs with self-focusing kerr-type nonlinearity. Phys. Rev. A **77**, 033821 (2008) https://doi.org/10.1103/PhysRevA.77.033821

[17] Radosavljević, A., Radovanović, J., Milanović, V.: Optimization of cubic gan/algan quantum well-based structures for intersubband absorption in the infrared spectral range. Solid State Communications **182**, 38–42 (2014)

[18] Radovanović, J., Milanović, V., Ikonić, Z., Indjin, D.: Application of the genetic algorithm to the optimized design of semimagnetic semiconductor-based spin-filters. Journal of Physics D: Applied Physics **40**(17), 5066 (2007) https://doi.org/10.1088/0022-3727/40/17/010

[19] Radovanović, J., Milanović, V., Ikonić, Z., Indjin, D.: Quantum-well profile optimization for maximal Stark effect and application to tunable infrared photodetectors. Journal of Applied Physics **91**(1), 525–527 (2002)

[20] Radovanović, J., Indjin, D., Milanović, V., Ikonić, Z.: Resonant intersubband harmonic generation in asymmetric bragg-confined quantum wells. Solid State

Communications **110**(6), 339–343 (1999)

[21] Vukovic, N., Radovanovic, J., Milanovic, V.: Enhanced modeling of band non-parabolicity with application to a mid-ir quantum cascade laser structure. Physica Scripta **2014**(T162), 014014 (2014) https://doi.org/10.1088/0031-8949/2014/T162/014014

[22] Pereira Jr, M.F., Henneberger, K.: Microscopic theory for the optical properties of coulomb-correlated semiconductors. physica status solidi (b) **206**(1), 477–491 (1998)

[23] Pereira, M.F.: Analytical expressions for numerical characterization of semiconductors per comparison with luminescence. Materials **11**(1) (2018) https://doi.org/10.3390/ma11010002

[24] Oriaku, C.I., Pereira, M.F.: Analytical solutions for semiconductor luminescence including coulomb correlations with applications to dilute bismides. J. Opt. Soc. Am. B **34**(2), 321–328 (2017) https://doi.org/10.1364/JOSAB.34.000321

[25] Wacker, A.: Semiconductor superlattices: a model system for nonlinear transport. Physics Reports **357**(1), 1–111 (2002)

[26] Wacker, A., Lindskog, M., Winge, D.O.: Nonequilibrium green's function model for simulation of quantum cascade laser devices under operating conditions. IEEE Journal of Selected Topics in Quantum Electronics **19**(5), 1–11 (2013) https://doi.org/10.1109/JSTQE.2013.2239613

[27] Pereira, M., Zubelli, J., Winge, D., Wacker, A., Rodrigues, A., Anfertev, V., Vaks, V.: Theory and measurements of harmonic generation in semiconductor superlattices with applications in the 100 ghz to 1 thz range. Physical Review B **96**(4), 045306 (2017)

[28] Pereira, M., Anfertev, V., Shevchenko, Y., Vaks, V.: Giant controllable gigahertz to terahertz nonlinearities in superlattices. Scientific Reports **10**(1), 15950 (2020)

[29] Vaks, V., Anfertev, V., Chernyaeva, M., Domracheva, E., Yablokov, A., Maslennikova, A., Zhelesnyak, A., Baranov, A., Schevchenko, Y., Pereira, M.F.: Sensing nitriles with thz spectroscopy of urine vapours from cancers patients subject to chemotherapy. Scientific Reports **12**(1), 18117 (2022)

[30] Pereira, M.F., Anfertev, V.A., Zubelli, J.P., Vaks, V.L.: Terahertz generation by gigahertz multiplication in superlattices. Journal of Nanophotonics **11**(4), 046022–046022 (2017)

[31] Apostolakis, A., Pereira, M.F.: Potential and limits of superlattice multipliers coupled to different input power sources. Journal of Nanophotonics **13**(3), 036017–036017 (2019)

[32] Apostolakis, A., Pereira, M.F.: Superlattice nonlinearities for gigahertz-terahertz generation in harmonic multipliers. Nanophotonics **9**(12), 3941–3952 (2020)

[33] Apostolakis, A., Pereira, M.F.: Controlling the harmonic conversion efficiency in semiconductor superlattices by interface roughness design. AIP Advances **9**(1) (2019)

[34] Pereira, M.F.: Harmonic generation in biased semiconductor superlattices. Nanomaterials **12**(9), 1504 (2022)

[35] Apostolakis, A., Awodele, M.K., Alekseev, K.N., Kusmartsev, F.V., Balanov, A.G.: Nonlinear dynamics and band transport in a superlattice driven by a plane wave. Phys. Rev. E **95**, 062203 (2017)

[36] Zhao, X., McKenzie, A.F., Munro, C.W., Hill, K.J., Kim, D., Bayliss, S.L., Gerrard, N.D., MacLaren, D.A., Hogg, R.A.: Large-area 2d selective area growth for photonic crystal surface emitting lasers. Journal of Crystal Growth **603**, 127036 (2023)

[37] Kyaw, A.S.M., Kim, D.-H., Butler, I.M., Nishi, K., Takemasa, K., Sugawara, M., Childs, D.T.D., Hogg, R.A.: Extreme temperature operation for broad bandwidth quantum-dot based superluminescent diodes. Applied Physics Letters **122**(3), 031104 (2023)

[38] McKenzie, A.F., Kyaw, A.M., Gerrard, N.D., MacLaren, D.A., Hogg, R.A.: Kinetic influences on void formation in epitaxially regrown gaas-based pcsels. Journal of Crystal Growth **602**, 126969 (2023)

[39] Apostolakis, A., Balanov, A.G., Kusmartsev, F.V., Alekseev, K.N.: Beyond the ordinary acoustoelectric effect: Superluminal phenomena in the acoustic realm and phonon-mediated bloch gain. Phys. Rev. B **106**, 014312 (2022)

[40] Zafar, H., Pereira, M.F., Kennedy, K.L., Anjum, D.H.: Fabrication-tolerant and cmos-compatible polarization splitter and rotator based on a compact bent-tapered directional coupler. AIP Advances **10**(12) (2020)

[41] Zafar, H., Zhai, Y., Villegas, J.E., Ravaux, F., Kennedy, K.L., Pereira, M.F., Rasras, M., Shamim, A., Anjum, D.H.: Compact broadband (o, e, s, c, l & u bands) silicon te-pass polarizer based on ridge waveguide adiabatic s-bends. Optics Express **30**(6), 10087–10095 (2022)

[42] Zafar, H., Paredes, B., Taha, I., Villegas, J.E., Rasras, M., Pereira, M.F.: Compact and broadband adiabatically bent superlattice-waveguides with negligible insertion loss and ultra-low crosstalk. IEEE Journal of Selected Topics in Quantum Electronics **29**(6: Photonic Signal Processing), 1–9 (2023)

[43] Zafar, H., Paredes, B., Villegas, J., Rasras, M., Pereira, M.F.: O-band te-and

tm-mode densely packed adiabatically bent waveguide arrays on the silicon-on-insulator platform. Optics Express **31**(13), 21389–21398 (2023)

[44] Razeghi, M., Lu, Q.Y., Bandyopadhyay, N., Zhou, W., Heydari, D., Bai, Y., Slivken, S.: Quantum cascade lasers: from tool to product. Opt. Express **23**(7), 8462–8475 (2015)

[45] Slivken, S., Razeghi, M.: Room temperature, continuous wave quantum cascade laser grown directly on a si wafer. IEEE Journal of Quantum Electronics **59**(4), 1–6 (2023)

[46] Slivken, S., Razeghi, M.: High power, room temperature inp-based quantum cascade laser grown on si. IEEE Journal of Quantum Electronics **58**(6), 1–6 (2022)

[47] Slivken, S., Shrestha, N., Razeghi, M.: High power mid-infrared quantum cascade lasers grown on si. Photonics **9**(9) (2022)

[48] Franckié, M., Faist, J.: Bayesian optimization of terahertz quantum cascade lasers. Phys. Rev. Appl. **13**, 034025 (2020)

[49] Meng, B., Hinkov, B., Biavan, N.M.L., Hoang, H.T., Lefebvre, D., Hugues, M., Stark, D., Franckié, M., Torres-Pardo, A., Tamayo-Arriola, J., Bajo, M.M., Hierro, A., Strasser, G., Faist, J., Chauveau, J.M.: Terahertz intersubband electroluminescence from nonpolar m-plane zno quantum cascade structures. ACS Photonics **8**(1), 343–349 (2021)

[50] Franckié, M., Bosco, L., Beck, M., Bonzon, C., Mavrona, E., Scalari, G., Wacker, A., Faist, J.: Two-well quantum cascade laser optimization by non-equilibrium Green's function modelling. Applied Physics Letters **112**(2), 021104 (2018)

[51] Winge, D.O., Franckié, M., Verdozzi, C., Wacker, A., Pereira, M.F.: Simple electron-electron scattering in non-equilibrium green's function simulations. In: Journal of Physics: Conference Series, vol. 696, p. 012013 (2016). IOP Publishing

[52] Clerici, M., Peccianti, M., Schmidt, B.E., Caspani, L., Shalaby, M., Giguere, M., Lotti, A., Couairon, A., Légaré, F., Ozaki, T., *et al.*: Wavelength scaling of terahertz generation by gas ionization. Physical Review Letters **110**(25), 253901 (2013)

[53] Vaks, V.: High-precise spectrometry of the terahertz frequency range: The methods, approaches and applications. J Infrared Milli Terahz Waves **33**, 43–53 (2012)

[54] Kozin, I.N., Polyansky, O.L., Pripolzin, S.I., Vaks, V.L.: Millimeter-wave transitions of h2se. Journal of Molecular Spectroscopy **156**(2), 504–506 (1992)

[55] Pankratov, A.L., Vaks, V.L., Koshelets, V.P.: Spectral properties of phase-locked flux flow oscillator. Journal of Applied Physics **102**(6), 063912 (2007)

[56] Koshelets, V.P., Ermakov, A.B., Filippenko, L.V., Kinev, N.V., Kiselev, O.S., Torgashin, M.Y., Lange, A., Lange, G., Pripolzin, S.I., Vaks, V.L.: Superconducting integrated THz receivers: development and applications. In: Zhang, C., Zhang, X.-C., Siegel, P.H., He, L., Shi, S.-C. (eds.) Infrared, Millimeter Wave, and Terahertz Technologies, vol. 7854, p. 78540. SPIE, ??? (2010). https://doi.org/10.1117/12.868916 . International Society for Optics and Photonics

[57] Vaks, D.E.G.C.M.B.e.a. V.L.: Methods and approaches of high resolution spectroscopy for analytical applications. Opt Quant Electron **49**, 239 (2016) https://doi.org/10.1007/s11082-017-1076-6

[58] Vaks, V.L., Domracheva, E.G., Chernyaeva, M.B., Pripolzin, S.I., Anfertev, V.A., Yablokov, A.A., Lukyanenko, I.A., Sheikov, Y.V.: High-resolution terahertz spectroscopy for investigation of energetic materials during their thermal decomposition. IEEE Transactions on Terahertz Science and Technology **11**(4), 443–453 (2021) https://doi.org/10.1109/TTHZ.2021.3074030

[59] Lykina, A.A., Anfertev, V.A., Domracheva, E.G., Chernyaeva, M.B., Kononova, Y.A., Toporova, Y., Korolev, D.V., Smolyanskaya, O.A., Vaks, V.L.: Terahertz high-resolution spectroscopy of thermal decomposition gas products of diabetic and non-diabetic blood plasma and kidney tissue pellets. Journal of Biomedical Optics **26**(4), 043008 (2021) https://doi.org/10.1117/1.JBO.26.4.043008

[60] Abina, A., Koro°ec, T., Puc, U., Jazbin°ek, M., Zidan°ek, A.: Urinary metabolic biomarker profiling for cancer diagnosis by terahertz spectroscopy: Review and perspective. Photonics **10**(9) (2023) https://doi.org/10.3390/photonics10091051

[61] Winnerl, S., Schomburg, E., Brandl, S., Kus, O., Renk, K., Wanke, M., Allen, S., Ignatov, A., Ustinov, V., Zhukov, A., *et al.*: Frequency doubling and tripling of terahertz radiation in a gaas/alas superlattice due to frequency modulation of bloch oscillations. Applied Physics Letters **77**(9), 1259–1261 (2000)

[62] Waschke, C., Roskos, H.G., Schwedler, R., Leo, K., Kurz, H., Köhler, K.: Coherent submillimeter-wave emission from bloch oscillations in a semiconductor superlattice. Physical review letters **70**(21), 3319 (1993)

[63] Zhang, W.L.Z.J.e.a. B.: Control of magnetic anisotropy by orbital hybridization with charge transfer in (la0.67sr0.33mno3)n/(srtio3)n superlattice. NPG Asia Mater **10**, 931–942 (2018)

[64] Isakovic, A.F., Berezovsky, J., Crowell, P.A., Chen, L.C., Carr, D.M., Schultz, B.D., Palmstrom, C.J.: Control of magnetic anisotropy in fesub 1minusxcosub x films on vicinal gaas and scsub 1minusyersub yas surfaces. Journal of Applied Physics **89**(11) (2001) https://doi.org/10.1063/1.1355320

[65] Alkhidir, T., Gater, M.A.J.D.L., Alpha, C., Isakovic, A.F.: Detection of some amino acids with modulation-doped and surface-nanoengineered gaas schottky p-i-n diodes. J. Vac. Sci. Technol. B **38**, 054002 (2020)

[66] Zhou, J., Liu, Y., Tang, J., Tang, W.: Surface ligands engineering of semiconductor quantum dots for chemosensory and biological applications. Materials Today **20**(7), 360–376 (2017) https://doi.org/10.1016/j.mattod.2017.02.006

[67] Tsu, R., Esaki, L.: Tunneling in a finite superlattice. Applied Physics Letters **22**(11), 562–564 (1973)

[68] Pereira, M.F., Apostolakis, A.: Combined structural and voltage control of giant nonlinearities in semiconductor superlattices. Nanomaterials **11**(5), 1287 (2021)

[69] Venetis, J.: An analytic exact form of the unit step function. Mathematics and Statistics **2**, 235–237 (2014)

[70] Schevchenko, Y., Apostolakis, A., Pereira, M.F.: Recent advances in superlattice frequency multipliers. Terahertz (THz), Mid Infrared (MIR) and Near Infrared (NIR) Technologies for Protection of Critical Infrastructures Against Explosives and CBRN, 101–116 (2021).